\documentclass[12pt]{article}
\usepackage{enumerate}
\usepackage{natbib}

\usepackage{amssymb}
\usepackage{amsfonts}
\usepackage{amsmath}
\usepackage{booktabs}
\usepackage{algorithm}
\usepackage{algpseudocode}
\usepackage{xcolor}
\usepackage{color}
\usepackage{colortbl}
\usepackage{multirow}
\usepackage{lscape}
\usepackage[font=small,labelfont=bf,tableposition=top]{caption}
\usepackage[subrefformat=parens]{subcaption}
\usepackage{graphicx}
\usepackage{float}
\usepackage{bm}
\usepackage{afterpage}
\usepackage{mathtools}
\usepackage{mathabx}
\usepackage[hyphens,spaces,obeyspaces]{url}
\usepackage{enumitem}
\usepackage{threeparttable}

\newcommand{\blind}{1}

\DeclareMathOperator*{\argmax}{arg\,max}

\newtheorem{remark}{Remark}

\newtheorem{theorem}{Theorem}

\newtheorem{definition}{Definition}
\newtheorem{corol}{Corollary}

\begin{document}

\newcommand{\grafml}{\mathcal{G}}  
\newcommand{\graph}{G}
\newcommand{\vertice}{V}
\newcommand{\edge}{E}

\newcommand{\data}{Z}
\newcommand{\datamat}{{\bf Z}}
\newcommand{\samplespace}{\mathcal{Z}}
\newcommand{\real}{{z}}  

\newcommand{\cova}{X}  
\newcommand{\resp}{Y}  
\newcommand{\mean}{W}
\newcommand{\covamat}{{\bf X}}
\newcommand{\respvec}{{\bf Y}}
\newcommand{\meanvec}{{\bf W}}
\newcommand{\creal}{x}  
\newcommand{\rreal}{y}  
\newcommand{\mreal}{w}

\newcommand{\intpar}{{J}}  
\newcommand{\extpar}{{H}}  
\newcommand{\logpar}{{\beta}}  
\newcommand{\dual}{{\eta}}
\newcommand{\sactive}{\mathcal{A}}
\newcommand{\sinactive}{\mathcal{I}}
\newcommand{\sexc}{\mathcal{S}}  

\newcommand{\ds}{\displaystyle}
\newcommand{\supp}{{\mathrm{supp}}}
\newcommand{\tr}{{\mathrm{tr}}}
\newcommand{\idct}{\mathbb{I}}  
\newcommand{\matfml}{\mathbb{S}_{0}^{p}}

\newcommand{\Var}{{\mbox{Var}}}
\newcommand{\Cov}{{\mbox{Cov}}}
\renewcommand{\vec}{\mathrm{vec}}

\newcommand{\cond}{{\psi}}  
\newcommand{\pl}{{L}}  
\newcommand{\npl}{{\mathcal{L}}}  
\newcommand{\ncl}{{\ell}}  
\newcommand{\gra}{{\bf G}}
\newcommand{\hes}{{\bf H}}
\newcommand{\tho}{{\bf U}}  

\newcommand{\mxeg}{{B}}
\newcommand{\mneg}{{b}}
\newcommand{\mxsg}{{\mu}}
\newcommand{\mnhs}{{m}}
\newcommand{\ws}{{\lambda}}  
\newcommand{\ms}{{\gamma}}  

\newcommand{\empn}{n^{\textup{emp}}}
\newcommand{\algn}{n^{\textup{alg}}}

\newcommand{\iter}{k}  
\newcommand{\dg}{d}  
\newcommand{\exc}{s}  

\newcommand{\spdspace}{\mathcal{S}^p_{+}}
\newcommand{\distas}[1]{\mathbin{\overset{#1}{\kern\z@\sim}}}

\renewcommand{\algorithmicrequire}{\textbf{Input:}}
\renewcommand{\algorithmicensure}{\textbf{Output:}}

\newcommand{\gap}{c_0}
\newcommand{\tech}{\nu}

\def\spacingset#1{\renewcommand{\baselinestretch}%
{#1}\small\normalsize} \spacingset{1}


\if1\blind
{
  \title{{\bf Reconstruct Ising Model with Global Optimality via SLIDE}\thanks{We acknowledge the Editor, Associate Editor, and three referees for their insightful and valuable comments that have helped improve the quality of this paper and the reproducibility materials.}}
  \author{{Xuanyu Chen\textsuperscript{$\dag$}, Jin Zhu\textsuperscript{$\dag$}, Junxian Zhu\thanks{The first three authors contributed equally to this paper and are listed in alphabetical order.}, Xueqin Wang\thanks{Wang's research is partially supported by the National Natural Science Foundation of China (12231017), the National Key R\&D Program of China (2022YFA1003803), and the National Natural Science Foundation of China (72171216, 71921001, and 71991474).} and Heping Zhang\thanks{Zhang's research is partially supported by the U.S. National Institutes of Health (R01HG010171 and R01MH116527) and NSF (DMS-2112711). }} \\
  {University of Science and Technology of China} \\ 
  {Sun Yat-Sen University, University of Michigan} \\
  {London School of Economics and Political Science} \\ 
  {University of Birmingham, National University of Singapore} \\
  {Yale University}}
  \date{}
  \maketitle
} \fi

\if0\blind
{
  \bigskip
  \bigskip
  \bigskip
  \begin{center}
    {\LARGE\bf {Reconstruct Ising Model with Global Optimality via SLIDE}}
\end{center}
  \medskip
} \fi

\bigskip
\begin{abstract}
  The reconstruction of interaction networks between random events is a critical problem arising from statistical physics and politics, sociology, biology, psychology, and beyond. The Ising model lays the foundation for this reconstruction process, but finding the underlying Ising model from the least amount of observed samples in a computationally efficient manner has been historically challenging for half a century. Using sparsity learning, {we present an approach named SLIDE whose sample complexity is globally optimal.} Furthermore, an algorithm is developed to give a statistically consistent solution of SLIDE in polynomial time with high probability. {On extensive benchmarked cases, the SLIDE approach demonstrates dominant performance in reconstructing underlying Ising models, confirming its superior statistical properties.} The application on the U.S. senators voting in the six congresses reveals that both the Republicans and Democrats noticeably assemble in each congress; interestingly, the assembling of Democrats is particularly pronounced in the latest congress.
\end{abstract}

\vfill

\newpage
\spacingset{1.82} 

\section{Introduction}\label{sec:intro}
In contemporary applications, the Ising model serves as a prevalent tool for characterizing the pairwise interactions between binary variables in a network system. In this network, each node represents a binary variable, the presence of an edge between any two nodes indicates that the corresponding variables have an interaction. Reconstructing the corresponding graph of the Ising model is critical to the studies of interaction in a broad spectrum of complex systems in biology, neuroscience, ecology, economics, political science, image analysis \citep{coccoNeuronalCouplingsRetinal2009, weistuchMetabolismModulatesNetwork2021, nobleSpatialPatternsTree2018, battistonComplexityTheoryFinancial2016, banerjee2008model, saremiHierarchicalModelNatural2013}. While many methods have been proposed during the past half century for reconstructing Ising models, finding the most sample-efficient estimate in a computationally efficient manner such as in a polynomial time remains an open problem. Given the profound importance of the Ising model, especially in the era of social media, there is an urgent need to solve this open problem. 

\subsection{Literature Review}
Reconstructing Ising models by maximizing the likelihood function is considered to be the gold standard, although it is computationally intractable since the evaluation of the partition function involves the sum of $2^p$ terms \citep{nguyen2017inverse}, where $p$ is the number of nodes in the Ising model. Many techniques have been developed to approximate the likelihood function. For instance, it is suggested that Monte Carlo methods can be used to approximate the partition function \citep{geyer1994convergence,miasojedow2018sparse}, and then the approximating likelihood function is maximized. Although this approach is asymptotically exact, it takes an exponential number of runs to achieve a predefined accuracy. Alternative methods that can avoid evaluating the partition function include the naive mean-field approximation and its advanced variants \citep{chow1968approximating, kappenBoltzmannMachineLearning1997, nguyenMeanFieldTheoryInverse2012}. These methods assume that the network is composed of some basic clusters, and when this assumption is met, they give the exact value of the likelihood function. As one increases the size of basic clusters, the methods apply to a broader class of networks, while the computational complexity also sharply grows. However, for networks beyond the applicable class, the effectiveness of such approximations remains uninvestigated. 

Instead of approximations, surrogates of the likelihood function also have been proposed to sidestep the computational cost. \citet{ravikumar2010high} proposed to node-wisely maximize the conditional likelihood. The product of conditionals, commonly referred to as pseudo-likelihood or composite likelihood, is often coupled with various regularizations and widely applied in the literature \citep{besag1975statistical, hofling2009estimation, jalali2011learning, xue2012nonconcave}. In \citet{vuffray2016interaction}, one proposed to use the interaction screening objective, which is closely related to the exponential loss of the logistic model. These methods are capable for general Ising models and can alleviate the prohibitive computation for their reconstruction. Typically, this strategy employs convex regularization, which may introduce bias and the subtle steps in tuning hyperparameters and selecting post-inference thresholds for theoretic guarantees \citep{lokhov2018optimal}. More importantly, we prefer methods that use the fewest samples to perfectly recover the underlying Ising model. However, every aforementioned method requires a sample size much larger than the necessary sample complexity derived in \citet{santhanam2012information}. The best result is the optimality along only a subset of the parameters \citep{lokhov2018optimal}. 

Graphical structure recovery resides in the realm of variable selection, for which the predominance of sparse learning approach has been demonstrated in a variety of statistical models \citep{raskutti2011minimax, saravandegeerEllPenalizedMaximum2013}. The sparse learning approach directly addresses the biasedness of the aforementioned regularization-based methods. Moreover, the approach may lead to stronger statistical properties, as they usually result in a smaller parameter space to search over. 
Finally, the development of statistical tools allows us to consider directly solving the sparse learning problem. For some well-studied models like (generalized) linear models, in recent years, statisticians have proposed several feasible approximate algorithms that solve the sparse regression under mild conditions \citep{huang2018constructive, zhu2020polynomial, zhang2021certifiably, zhu2021abess}. However, for Ising models, the potential benefits of the sparse learning approach have not yet been rigorously authenticated, and how to design a feasible algorithm that captures the essence of sparse learning for reconstructing the Ising model remains another meaningful problem. 

\subsection{Contributions}
In this paper, we propose a new method named SLIDE (short for \underline{s}parsity \underline{l}earning for \underline{I}sing mo\underline{d}el r\underline{e}construction), which considers a direct but largely unexplored approach --- sparsity learning. By using pseudo-likelihood as the objective function, we derive the sample complexity of SLIDE. {Two appealing theoretical advantages are demonstrated by the derived sample complexity. First, SLIDE's sample complexity is optimal with respect to all structural parameters that are of interest, which we refer to as \textit{global optimality}. Second, to the best of our knowledge, the sample complexity of SLIDE is lower than that of any existing method.} Building upon this theoretical foundation, we proceed to devise a tractable algorithm for identifying the true graphical structure of an Ising model, with provable computational scalability and statistical consistency. Besides, we establish the dominance of our proposal over other state-of-the-art methods via simulations under both asymptotic and non-asymptotic regimes. Finally, applying our method to a real-world dataset uncovers intriguing patterns in the voting behavior of U.S. senators.

\section{Preliminaries}
\subsection{Ising Model and its Reconstruction}\label{sec:pre:ising-model}

Consider a graph $G=(V,E)$ with $p$ nodes where $V= \{1,...,p\}$ is the node set and $E \subseteq V\times V$ is the undirected edge set. Here, a node $i \in V$ is associated with a binary random variable $z_i \in \{-1,+1\}$. An edge $(i,j) \in E$ is associated with a non-zero real parameter $\intpar_{ij} = \intpar_{ji} \neq 0$ that is called 
the interaction between node $i$ and $j$. In the language of statistics, the Ising model\footnote{For simplicity, we consider Ising models with zero external magnetic fields, i.e., $J_{ii} = 0$ for $i=1, \ldots, p$.} is a joint distribution of $(z_1, \ldots, z_p)$ whose probability mass function reads
\begin{align*}
	P_{J, G}(z) = \frac{1}{\Phi(\intpar)}\exp \Big(\sum_{(i, j)\in E}\intpar_{ij}\real_{i}\real_{j}\Big),
\end{align*}
where $J$ is a $p$-by-$p$ symmetric matrix and $\Phi(\intpar)$ is the partition function that ensures $\sum_{z\in \{-1,1\}^p}P_{J, G}(z) = 1$. Throughout the paper, we focus on models whose underlying distributions $P_{J,G}$ belong to the following family of distributions, which is widely used in the literature \citep{lokhov2018optimal}.
\begin{definition}[Family of distributions, \citet{santhanam2012information}]\label{def:interest-ising-model}
For a pair of positive constants~$(\ws,\ms)$, $\grafml_{p,\dg}(\ws,\ms)$ is the family of distributions $P_{\intpar, \graph}$ satisfying (i) the graph $\graph\in \grafml_{p,\dg}$ and (ii) the parameter matrix $\intpar\in \intpar_{\ws,\ms}(\graph)$. Here, $\grafml_{p,\dg}$ is the class of graphs with $p$ nodes and bounded node degree $\dg$, and $\intpar_{\ws,\ms}(\graph)$ is the set of zero-diagonal symmetric matrices that satisfy: $\intpar_{ij}\neq 0$ if and only if $(i,j)\in \edge$, the minimum signal $\min\limits_{(i,j)\in\edge}|\intpar_{ij}| \geq \ws$, and the maximum neighborhood weight $\max\limits_{i\in \vertice} \sum\limits_{j=1}^{p}|\intpar_{ij}| \leq \ms$.
\end{definition}
According to (ii) in Definition~\ref{def:interest-ising-model}, we can rewrite the probability mass function as
\begin{align*}
	P_{J}(z) = \frac{1}{\Phi(\intpar)}\exp \Big(\sum_{i < j}\intpar_{ij}\real_{i}\real_{j}\Big)
\end{align*}
since $\intpar_{ij}=0$ implies $(i, j)\notin E$. Note that, we occasionally omit the subscript $\intpar$ in $P_{\intpar}$ for simplicity when there is no ambiguity; besides, we define the support set of matrix $\intpar$ as $\supp(\intpar)$. 

Suppose $\intpar^*$ is the unknown true parameters that characterizes variables' interactions, 
Ising model reconstruction aims to recover $\intpar^*$ via $n$ observed samples $z^{(1)}, \ldots, z^{(n)} \in \mathbb{R}^p$ that are independently sampled from the Ising model $P_{\intpar^*}$. Specifically, in this paper, we focus on this problem when the underlying graph only has a relatively small number of edges, i.e., the vast majority of the $\intpar_{ij}$'s are exactly zero. This sparsity assumption is reasonable as it gives better interpretability of the model. In the meanwhile, it appears to be a necessary assumption when the dimension $p$ is comparable to or larger than the sample size $n$ (the so-called high-dimensional regime).

\subsection{Globally Optimal Sample Complexity Regime}\label{sec:pre:sample-complexity}
\textbf{Background.} For the reconstruction of Ising models, the sample complexity of an estimator refers to the minimum sample size such that, with a high probability, this estimator can recover any underlying Ising model belongs to the family of distributions introduced in Definition~\ref{def:interest-ising-model}. For Ising models satisfying Definition~\ref{def:interest-ising-model}, \citet{santhanam2012information} derived the information-theoretic lower bound of the sample size with respect to four structural parameters: maximum neighborhood weight $\ms$, maximum node degree $\dg$, number of nodes $p$, and minimum signal~$\lambda$. These parameters are of primary interest when studying the sample complexity. Specifically, \citet{santhanam2012information} showed that the information-theoretic lower bound is of order
\begin{align}\label{eq:lb}
    n_{l} = \Omega\left((1-\varepsilon)\max\left\{\frac{1}{\ws^2},\frac{e^\ms}{\ws d}, d\right\}\log p\right).
\end{align}
Moreover, to show the tightness of the above lower bound, \citet{santhanam2012information} provided an attainable sample complexity, by analyzing the sample complexity of a projection-type graphical structure estimator. To be specific, this estimator returns the graphical structure of the model whose second-order moments best match their empirical counterparts. Such an estimator is computationally intractable, but it offers a sample complexity that is attainable (at least theoretically) by an estimator:
\begin{equation}\label{eq:ub}
\begin{aligned}
n_{{u}}
= O\left(\frac{\ms^2}{\ws^4}e^{4\ms}\log \left(\frac{p}{\varepsilon}\right)\right).
\end{aligned}
\end{equation}
The optimal sample size $n_{\text{opt}}$ lies within $n_l$ and $n_u$ yet its specific form remains unknown. From $n_l$ and $n_u$, we know that the optimal sample complexity $n_{\text{opt}}$ has the following dependencies on $(\ws, \ms, \dg, p)$: $\ds n_{\text{opt}} \propto \log p$, $ n_{\text{opt}} \propto \ws^{-2}$, $\ds n_{\text{opt}} \propto \dg^{r}$	and $\ds n_{\text{opt}} \propto e^{c\ms}$ where $r$ and $c$ are two constants satisfying $r\in [1,2]$ and $c\in [1,4]$.

\noindent\textbf{Globally optimal sample complexity.} We introduce the ``globally optimal'' regime which is defined as:
\begin{equation}\label{eq:global-optimality}
	n \propto \frac{1}{\ws^2}, \; n \propto \log(p), \; n \propto \exp(c\ms), \; n \propto d,
\end{equation}
for any $c \in [1, 4]$. An estimator whose sample complexity satisfies~\eqref{eq:global-optimality} is called a globally optimal estimator. The regime~\eqref{eq:global-optimality} integrates insights from both the information-theoretic lower bounds~\eqref{eq:lb} and the attainable sample complexities~\eqref{eq:ub} derived by \citet{santhanam2012information}. By defining the regime as the lower and upper limits for each critical structural parameter, this regime comprehensively covers the entire range of these parameters. The rationality of the globally optimal regime~\eqref{eq:global-optimality} is twofold.
\begin{itemize}[leftmargin=*]
\item \textbf{Tightness.} To the best of our knowledge, no existing algorithm has a lower sample complexity than our upper boundary, showing this regime as the tightest one that contains the optimal rate in the literature. Additionally, this regime incorporates the conjecture by \citet{santhanam2012information} that the information-theoretic lower bound for $d$ is tight, further reinforcing the tightness of the defined regime.  
\item \textbf{Generality.} The global optimality regime captures all relevant structural parameters. In contrast, the optimal regime proposed by \citet{lokhov2018optimal}, i.e., 
$$n \propto \frac{1}{\ws^{2}},\; n \propto \log(p),\; n \propto \exp(c\ms),$$
overlooked the dependency on the maximum degree $d$. Their regime implicitly assumes $d = o(e^{\gamma})$, where the effect of $d$ in both of~\eqref{eq:lb} and~\eqref{eq:ub} can be ignored, and they merely focus on the effect of $\ms$. However, this assumption may not hold for general Ising models and a counterexample is constructed in Section~\ref{sec:experiment-on-d}. By removing the implicit assumption, the globally optimal regime serves as a more comprehensive extension. 
\end{itemize} 

\section{{SLIDE and its Sample Complexity}}
In Section~\ref{sec:simple}, we describe a sparse learning framework for reconstructing the network associated with the Ising model. Impressively, as we will discuss in Section~\ref{sec:thc_complexity}, the theoretical support for this framework shows that it {is a globally optimal estimator}. 
\subsection{Sparse Learning for Ising Model Reconstruction}\label{sec:simple}
We begin by introducing notations for the presentation of our framework.
Let $\intpar_{i}$ be the $i$-th column of the interaction matrix $\intpar$, then for each node~$i$ in this network, we define logarithmic pseudo-likelihood (PL) as 
\begin{align*}
	\npl_n(\intpar_i) \coloneqq & \sum_{r=1}^{n}\log P(\real_{i}^{(r)}|\ \real_{1}^{(r)}, \ldots, \real_{i-1}^{(r)}, \real_{i+1}^{(r)}, \ldots,\real_{p}^{(r)}).
\end{align*}
To put it simply, when the interaction between node $i$ and the others is $\intpar_i$, the PL measures the probability of observing node~$i$'s collected samples given the other nodes' samples. To rephrase, $\npl_n(\intpar_i)$ measures the reliability of $\intpar_i$ given the observed data. Having a greater value for $\npl_n(\intpar_i)$ indicates that $\intpar_i$ adequately explains the variance of node~$i$.
Next, we define the number of neighbors of node $i$, $\mathcal{N}(\intpar_i) \coloneqq \#\{\intpar_{ij}\neq 0\ |\ j\neq i\}$. Indeed, $\mathcal{N}(\intpar_i)$ naturally characterizes the complexity of the local system of node~$i$. 

{Our framework is expressed as a compromise between local system accuracy $\npl_n(\intpar_i)$ and local system complexity $\mathcal{N}(\intpar_i)$, which is defined as: 
\begin{equation}\label{eq:opt_constrain2}
	\widehat{\intpar}_i \leftarrow \arg\max_{\intpar_i} \; \npl_n(\intpar_i) \quad \textup{s.t. } \mathcal{N}(\intpar_i) \leq \dg, \min_{(i, j)\in \edge} |J_{ij}| > \ws, \sum_j |J_{ij}| < \ms  \ (\textup{for } i = 1, \ldots, p).
\end{equation}
Note that the constraints involving $\ms$ and $\ws$ are introduced to fulfill Definition~\ref{def:interest-ising-model}. The term for this approach that we've coined is \underline{s}parse \underline{l}earning for \underline{I}sing mo\underline{d}el r\underline{e}construction (SLIDE). Concatenating $\widehat{J} \coloneqq (\widehat{\intpar}_1, \ldots, \widehat{\intpar}_p)$, the solutions of \eqref{eq:opt_constrain2}, and then performing an additional matrix symmetrization step, i.e., $\widehat{J} \leftarrow \frac{1}{2}(\widehat{J} + \widehat{J}^\top)$, completes the reconstruction of an Ising model.}
\begin{remark}
The advantage of SLIDE is two-fold. First, it meets the need for reconstructing networks by selecting the neighbors of each node. Once neighbors of every node are correctly found, the network structure must be identical to the true one. Particularly, the local system complexity constraint conducts the selection of neighbors straightforwardly,  which is the key reason for the superiority of SLIDE. In contrast, the other indirect constraints \citep{lokhov2018optimal} would need higher sample complexity to implicitly select the neighbors. Second, it exploits the PL $\npl_n(\intpar_i)\; (i=1, \ldots, p)$ as a surrogate of the computationally intractable likelihood function. It has an explicit expression:
\begin{align*}
    \npl_n(\intpar_i) = -\sum_{r=1}^{n}\log\Big\{\exp\Big(-2\sum_{j:j\neq i}\intpar_{ij}\real_{i}^{(r)}\real_{j}^{(r)}\Big) + 1\Big\}, 
\end{align*}
and computing the PL has an $O(np)$ time complexity and thus can be efficiently computed. 
\end{remark}
\vspace*{-10pt}
{\begin{remark}
A careful reader may notice that~\eqref{eq:opt_constrain2} does not exploit the symmetry of $\intpar$ to learn the structure. On the one hand, this is because the solution of~\eqref{eq:opt_constrain2} already exactly recover the underlying graph structure (see Theorem~\ref{thm:sample_size_d}). On the other hand, in the asymptotic regime, accounting for the symmetry of \( \intpar \) does not offer additional benefits to the recovery but causes the computational process to be more complicated as it has to consider $O(p^2)$ parameters simultaneously. In the finite-sample regime, we acknowledge leveraging symmetry might improve numerical performance \citep{hofling2009estimation, miasojedow2018sparse}, and we develop a symmetry-aware SLIDE variant in the Supplementary Materials. 
\end{remark}}

\subsection{Sample Complexity of SLIDE}\label{sec:thc_complexity}

Allowing the quadruple $(\ws, \ms, \dg, p)$ to scale with the sample size $n$, the following theorem gives the theoretic sample complexity of SLIDE to successfully recover the underlying graphical structure.

{\begin{theorem}\label{thm:sample_size_d}
Assume that $e^{-2\ms}\ws^2 \leq 8$. 
Suppose that $z^{(1)}, \ldots, z^{(n)}$ are \textit{i.i.d.} drawn from some $P_{\intpar^*}\in\grafml_{p,\dg}(\ws,\ms)$.
For any constant $\varepsilon\in(0, 1)$, if the sample size $n$ satisfies
$$
	n\geq n^{*} = \frac{128e^{4\ms}}{\ws^2}\Bigg(3\dg\log p +\dg\log \Big(1+\frac{384\ms\dg e^{4\ms}}{\ws^2}\Big) + \log\frac{1}{\varepsilon}\Bigg),
$$
then we have $\ds P_{\intpar^*}\big( \supp(\widehat{J}) \neq \supp(\intpar^*) \big)\leq \varepsilon$.
\end{theorem}
Our sample complexity $n^*$ can be presented more concisely (up to some logarithmic factors) as
$$
n\geq n^{*} = O\left(\frac{\dg}{\ws^2}e^{4\ms}\log \frac{p}{\varepsilon} \right).
$$
Clearly, the sample complexity of SLIDE lies in the globally optimal regime~\eqref{eq:global-optimality}, showing that SLIDE is a globally optimal estimator. Furthermore, to the best of our knowledge, along every axis of $(\ws, \ms, \dg, p)$, it matches or improves the theoretic sample complexities of all existing state-of-the-art algorithms listed in Table \ref{tab:sc-comparison2}. }

\begin{table}[h]
\begin{threeparttable}
\caption{{Dependency of the sample complexity of existing estimators on four structural parameters: maximum neighborhood weight $\ms$, maximum node degree $\dg$, number of nodes $p$, and minimum signal~$\ws$. MLE: maximum likelihood estimator. }}
\label{tab:sc-comparison2}
\centering
\vspace{-5pt}
{\small\begin{tabular}{l|cccc}
\toprule
Estimator   & $\ms$     & $\dg$         & $p$     & $\ws$      \\
\midrule
Projection-type estimator \citep{santhanam2012information}  & $e^{4\ms}$ & $d^2$ & $\log p$ & $\ws^{-2}$ \\
$\ell_1$-constrained nodewise regression \citep{rigollet2017high}    & $e^{8\ms}$   & $\dg^2$      & $\log p$ & $\ws^{-2}$ \\
$\ell_1$-regularized nodewise regression$^\dagger$ \citep{lokhov2018optimal}   & $e^{8\ms}$   & $\dg^4$      & $\log p$ & $\ws^{-2}$ \\
$\ell_1$-regularized interaction screening \citep{vuffray2016interaction}  & $e^{6\ms}$  & $\dg^4$      & $\log p$ & $\ws^{-2}$ \\
$\ell_1$-regularized log-interaction screening \citep{lokhov2018optimal} & $e^{12\ms}$  & $\dg^4$      & $\log p$ & $\ws^{-2}$ \\
$\ell_1$-regularized approximated MLE$^{*}$ \citep{miasojedow2018sparse} & $\geq e^{4\ms}$  & $\dg^2$      & $\log p$ & $\ws^{-2}$ \\
\midrule
\textbf{SLIDE (ours)}  & $e^{4\ms}$  & $\dg$      & $\log p$ & $\ws^{-2}$ \\
\bottomrule
\end{tabular}}
{\begin{tablenotes}[leftmargin=*]
\linespread{1}\footnotesize  
  \item[*] \citet{miasojedow2018sparse} does not explicitly demonstrate how sample complexity depends on $\ms$. In our analysis, we derive a dependency of $\geq e^{4\ms}$ based on their reported sample complexity under a specific coupling matrix configuration. For further details, please refer to the Supplementary Materials.
  \item[$\dagger$] The comparison to a similar estimator proposed by \citet{raskutti2011minimax} is discussed in the Supplementary Materials. 
\end{tablenotes}}
\end{threeparttable}
\end{table}

The SLIDE's sample complexity scales linearly in the maximum number of neighbors $\dg$, while other existing methods have theoretic sample complexities growing quadratically in $\dg$. {This is particularly important in practice, as systems in fields such as neuroscience and sociology often involve networks with a larger $d$ \citep{grabowskiOpinionFormationSocial2009, marinazzoInformationTransferCriticality2014}. In these domains, our method would possess superior performance compared to existing algorithms. Theoretically, our method allows the true sparse network to have a higher node degree.} Most notably, our method proves a conjecture in \citet{santhanam2012information} as its sample complexity scales linearly with $\dg$. We know that the optimal sample complexity scales with $\dg^r$, where $r \in [1, 2]$ is an unknown precise number conjectured to be $1$, according to the work of \citet{santhanam2012information}. By providing support for the claim that $r = 1$ for the SLIDE, our finding verifies the truth of this supposition.

For the maximum neighborhood weight $\ms$, the SLIDE also {lies in the globally optimal regime}. In the case of large $\ms$, the underlying Ising model is low-temperature, which is shown to be challenging to reconstruct \citep{montanariWhichGraphicalModels2009, nguyenMeanFieldTheoryInverse2012} and attracts keen attention from researchers \citep{decellePseudolikelihoodDecimationAlgorithm2014a, nguyenMeanFieldTheoryInverse2012}. Roughly speaking, the difficulty arises since in this case, most samples we collect would be of the network's ground states. To ensure a sufficient number of non-trivial samples, the sample size needs to grow exponentially as $\ms$ increases. Therefore, SLIDE has a greater potential for widespread use because it requires significantly fewer samples to reconstruct a low-temperature Ising model than existing approaches. Examples illustrating this advantage are provided in Figure~\ref{fig:temp-scale}. 

Last but not least, the sample complexity of SLIDE in terms of the number of nodes $p$ and the minimum interaction $\ws$ is a perfect match to the optimal sample complexity given by \citet{santhanam2012information}, i.e., it is proportional to $\log{p}$ and $\ws^{-2}$. For high-dimensional binary variables, where $p$ can scale exponentially with $n$, the former result is significant. The last implication means that no alternative structure recovery approach can use fewer samples than our proposal when the signal-to-noise ratio is relatively low. 

\section{Algorithm and its Properties}\label{sec:algo}

The time complexity grows exponentially with $p$ when using exhaustive enumeration to solve SLIDE. Therefore, to enhance the practical usability of SLIDE, we present an iterative algorithm in Section~\ref{sec:algo-decription}. The subsequent Section~\ref{sec:alg_theory} will exhibit both the computational and statistical properties of this algorithm.   

\subsection{Algorithmic Description}\label{sec:algo-decription}

Our algorithm mainly consists of three nested loops, as summarized in Algorithm~\ref{alg:abess-mcl}. In the innermost loop, our objective is to find the neighborhood of each fixed node $i$ when the number of neighbors $\dg$ is fixed. Regarding the middle loop, its purpose is to select the best neighborhood size $\dg$ for each fixed node $i$. Consequently, the middle loop returns the neighbor set for the fixed node $i$. Finally, the outer loop simply concatenates the results of the middle loop to reconstruct the entire graph. Below, we will provide more details about the inner and middle loops.

The inner loop employs the splicing technique proposed by~\citet{zhu2020polynomial}. The idea behind the \textit{splicing} technique is simple to grasp. For a given node $i$, if we have a current guess for its neighbors, we may improve the quality of our guess by including the nodes that are important to improve $\npl_n(\intpar_i)$ and eliminating the ones that have less important on $\npl_n(\intpar_i)$. Below, we provide a formal characterization of the importance. Suppose the neighborhood $\sactive\subseteq \vertice \setminus \{i\}$ and the interaction $\widehat{\intpar_i}=\argmax\limits_{\supp(\intpar_i) = {\sactive}} \npl_n(\intpar_i)$ are our current guess, we introduce two types of importance: (i) the \textit{backward importance} $\widehat{\intpar}_{ij}^2$ and (ii) the \textit{forward importance} $\widehat{\dual}_{j}^2$ where $\widehat{\dual}_{j} = \nabla_j\npl_n(\widehat{\intpar}_i)$.
\begin{remark}
	Backward importance can be interpreted as the price we pay for removing node $j \in \sactive$ from the current neighborhood $\sactive$, as one can show $C \widehat{\intpar}_{ij}^2 \approx \npl_n\big(\widehat{\intpar}_i\big) - \npl_n\big(\widehat{\intpar}_i |_{\sactive \setminus \{j\}} \big)$
	where $C$ is some universal constant and $\widehat{\intpar}_i |_{\sactive \setminus \{j\}}$ is obtained by setting the $j$th entry of $\widehat{\intpar}_i$ to zero. Analogously, forward importance can be interpreted as the gain of including node $j \in \sactive^c$ into $\sactive$ due to $C' \widehat{\dual}_{j}^2 \approx \max\limits_{t\in\mathbb{R}^{p-1},\ \supp(t) = \{j\}} \npl_n\big(\widehat{\intpar}_i + t\big) - \npl_n\big(\widehat{\intpar}_i\big)$, where $C'$ is some universal constant. 
\end{remark}
Furthermore, for any given $\exc\leq \dg$, define
\begin{equation}\label{eqn:splicing-set}
	\begin{split}
		\sexc_{\exc,1}&= \{j\in \sactive: \widehat{\intpar}^2_{ij} \text{ is one of the last $s$ backward importance}\},\\
		\sexc_{\exc,2}&= \{j\in \sactive^c: \widehat{\dual}^2_j \text{ is one of the top $s$ forward importance}\}.
	\end{split}
\end{equation}
Then, we splice $\sactive$ to create a new neighborhood structure by letting $\widetilde{\sactive} = (\sactive\setminus \sexc_{\exc,1}) \cup \sexc_{\exc,2}$. Setting $\widetilde{\intpar_i}=\argmax\limits_{\supp(\intpar_i) = \widetilde{\sactive}} \npl_n(\intpar_i)$, then $\widetilde{\mathcal{A}}$ is deemed to be a better guess than $\mathcal{A}$ if $\npl(\widetilde{\intpar}_i)>\npl_n(\widehat{\intpar}_i)$, then we shall update $(\sactive, \widehat{\intpar}_i)$ with $(\widetilde{\sactive}, \widetilde{\intpar}_i)$. The splicing technique is iteratively applied until the inner loop cannot find a better neighbor set. 

The main idea of middle loop is to determine the optimal neighbor size $d$ for the fixed node $i \in \vertice$ that guarantees a good balance between the system's accuracy and its complexity. {In the literature, the balance is generally measured by information criterion such as Akaike information criterion and Bayesian information criterion \citep{akaike1974new,schwarz1978estimating}. In this paper, we propose to employ a \textit{generalized information criterion} (GIC)}\footnote{{Several information criteria are referred to as GIC in the literature. We defer the discussion with them in the Supplementary Materials. }}. Specifically, given the interaction of the $i$-th node returned by the inner loop under the neighbor size $d_i$, denoted as $\widehat{\intpar}_{i, \dg_i}$, GIC assesses $\widehat{\intpar}_{i, \dg_i}$ by: 
$$\textup{GIC}(\widehat{\intpar}_{i, \dg_i}) \coloneqq \npl_{n}(\widehat{\intpar}_{i, \dg_i}) - d_i c_n \log p,$$ 
where the term $d_i c_n \log p $ encourages using fewer neighbors to explain the data so as to prevent overfitting caused by the incorrect inclusion of neighbors that weakly interact with node $i$. The proposed GIC is carefully designed to ensure consistent structure recovery. Specifically, $\log p$ is introduced to address overfitting in high-dimensional settings where $p$ is large.
In our algorithm, we set $c_n = \log\log n$ to ensure the sequence diverges slowly. To get the optimal number of neighbors, the middle loop selects a $d_i$ from $\{0, 1, \ldots, d_{\max}\}$ that maximizes $\textup{GIC}(\widehat{\intpar}_{i, \dg_i})$. 

\begin{remark}
	Two features of Algorithm~\ref{alg:abess-mcl} make it even more practically intriguing. First, the algorithm exploits the fact that the number of unknown best neighbors from which to choose is discrete. Selecting a discrete neighbor's number is more interpretable than choosing a continuous tuning parameter as the approaches in the literature. It is even more useful when there is a priori knowledge or expertise about the actual number of neighbors. For LASSO-type methods, this information has to be leveraged in a post-hoc manner. Furthermore, during iterations, the proposed algorithm estimates the interactions between nodes without shrinking them, leading to an unbiased estimate of the interactions. Consequently, our algorithm gives more accurate estimation of the interactions, especially in the small-sample regime, which can be witnessed by the smallest interaction estimation errors in Figure~\ref{fig:high-vs}. Finally, in most cases, it is also challenging to know a priori what the smallest interaction $\ws$ is. Notably, this procedure implicitly determines $\lambda$ in a data-driven way and gives a proper estimation for model complexity. 
\end{remark}

\begin{algorithm}[htbp]\caption{}\label{alg:abess-mcl}
	\begin{algorithmic}[1]
		\For{$i = 1,\ldots, p$} \algorithmiccomment{\textbf{Outer loop}: nodewise sparse learning}
		\State $\widehat{\intpar}_{i, 0} \leftarrow \mathbf{0}$, $\dual_{i, 0} \leftarrow \nabla\npl_n(\widehat{\intpar}_{i, 0})$.
		\For{{$\dg = 1, \ldots, \dg_{\max}\coloneqq \lceil\frac{n}{\log{p} \log\log{n}}\rceil$}} \algorithmiccomment{\textbf{Middle loop}: Searching the optimal $\dg$}
		\State $\iter \leftarrow -1$, $\sactive^{(0)} \leftarrow \supp(\widehat{J}_{i, d-1}) \cup \{ \argmax\limits_i (\dual_{i, d-1})^2 \}$.  \algorithmiccomment{Warm-start initialization}
		\Repeat \algorithmiccomment{\textbf{Inner loop}: iteratively employing the splicing}
		\State $\iter \leftarrow \iter + 1$, ${\intpar}_i^{(\iter)} \leftarrow \argmax\limits_{\supp(\intpar_i) = \sactive^{(\iter)}} \npl_n({\intpar_i}),\ \dual^{\iter} \leftarrow \nabla \npl_n({\intpar_i^{\iter}})\big|_{(\sactive^{\iter})^c}$.
		\For {{$\exc=1,\ldots, d$}} 
		\State Compute $\sexc_{\exc, 1}^{(\iter)} \ \text{and}\ \sexc_{\exc, 2}^{(\iter)}$ according to~\eqref{eqn:splicing-set}.
		\State $\widetilde{\sactive}^{(\iter)} \leftarrow (\sactive^{(\iter)}\backslash \sexc_{\exc, 1}^{(\iter)})\cup \sexc_{\exc, 2}^{(\iter)}$ and solve
		$\widetilde{\intpar}_i^{(\iter)} =\argmax\limits_{\supp(\intpar_i) = \widetilde{\sactive}^{(\iter)}} \npl_n({\intpar_i})$.
		\If {$\npl_n(\widetilde{\intpar}_i^{(\iter)}) - \npl_n(\intpar_i^{(\iter)}) > \sigma_\dg \coloneqq 0.01\frac{\dg \log p \log\log n}{n} $}
		\State $(\sactive^{\iter+1}, \intpar_i^{\iter+1}, \dual^{\iter+1}) \leftarrow (\widetilde{\sactive}^{(\iter)}, \widetilde{\intpar}_i^{(\iter)}, \nabla \npl_n({\widetilde\intpar_i^{\iter}})\big|_{(\sactive^{\iter})^c})$
		\EndIf
		\EndFor
		\Until{$\sactive^{\iter} = \sactive^{\iter - 1}$}
		\State $\widehat{\intpar}_{i, d} \leftarrow J_i^{\iter}, \dual_{i, d} \leftarrow \dual^{\iter}$.
		\algorithmiccomment{Solution of \textbf{Inner Loop}}
		\EndFor
		\State $\widehat{\dg} = \argmax\limits_{0\leq \dg\leq \dg_{\max}}\text{GIC}(\widehat{\intpar}_{i, d})$, $\widehat{\intpar}_i \leftarrow \widehat{\intpar}_{i, \widehat{\dg}}$.  \algorithmiccomment{Solution of \textbf{Middle Loop}}
		\EndFor
		\For{$i, j = 1, \ldots, p$} \algorithmiccomment{Symmetrize and threshold the solution of \textbf{Outer Loop}}
		\State $\widehat{\intpar}_{ij} \leftarrow \frac{1}{2}(\widehat{\intpar}_{ij} + \widehat{\intpar}_{ji}) \idct(\frac{1}{2}|\widehat{\intpar}_{ij} + \widehat{\intpar}_{ji}| \geq \tau)$ \algorithmiccomment{{$\tau = \ws/2$ if $\ws$ is known; otherwise $\tau = 0$.}}
		\EndFor
		\Ensure $\widehat{\intpar}$.
	\end{algorithmic}
\end{algorithm}
{We close this section with a discussion of some steps in Algorithm~\ref{alg:abess-mcl}.
\begin{itemize}[leftmargin=*]
	\vspace*{-8pt}
	\item \textbf{Initialization strategy.} In Step 4, the initialization of the inner loop sets the initial guess of $d$ neighbors as the union of (i) the previously estimated $d-1$ neighbors and (ii) the node with the highest forward importance. Notice that, the initial neighbor set $\sactive^{(0)}$ includes the previous estimated $d-1$ neighbors, i.e., $\supp(\widehat{J}_{i, d-1})$. This is a so-called warm-start initialization, a technique frequently employed in penalized regression to enhance empirical performance \citep[see, e.g.,][]{hastie2014glmnet}. Furthermore, using $\{ \argmax\limits_i (\dual_{i, d-1})^2 \}$ bears a similar idea of marginal variable screening methods \citep{fan2008sure}. To see this, consider the scenario where $d=1$ and $\sactive^{(0)} = \supp(\widehat{J}_{i, 0}) = \emptyset$. In this scenario, we have $\dual_{i, 0} = \left(\sum\limits_{r=1}^n \real^{(r)}_{1} \real^{(r)}_{i}, \ldots, \sum\limits_{r=1}^n \real^{(r)}_{i-1} \real^{(r)}_{i}, \sum\limits_{r=1}^n \real^{(r)}_{i+1} \real^{(r)}_{i}, \ldots, \sum\limits_{r=1}^n \real^{(r)}_{p} \real^{(r)}_{i} \right),$ which records the empirical covariance between the $j$-th node and the $i$-th node (where $i \neq j$). Therefore, $\sactive^{(0)}$ is the node with the largest magnitude of covariance with the $i$-th node.
	\vspace*{-30pt}
	\item \textbf{Leverage information of $\dg$ and $\ws$.} If the maximum degree $\dg$ is known, we can set $\dg_{\max} = \dg$. When the minimum signal $\ws$ is known, a post thresholding (see Step 20 in Algorithm~\ref{alg:abess-mcl}) can be applied on the estimated coupling matrix $\widehat{J}$ so as to remove the incorrectly identified edges that correspond to negligible estimated coupling.  Therefore, the thresholding procedure helps reduce falsely identified edges. In such cases, we suggest the threshold $\tau = \ws/2$ as in \citet{lokhov2018optimal}. When no knowledge on $\dg$ and $\ws$ is available, the theoretic analysis for Algorithm~\ref{alg:abess-mcl} suggests setting $\dg_{\max} = \lceil\frac{n}{\log{p} \log\log{n}}\rceil$ and $\tau = 0$ can guarantee the output enjoys statistical and computational properties.
	\vspace*{-8pt}
	\item \textbf{Threshold $\sigma_\dg$.} The $\sigma_\dg$ in Step 10 is introduced to reduce unnecessary splicing steps, when the model size $\dg$ is underestimated or the algorithm has already recovered the true graphical structure. Condition~\ref{con:threshold} for the theoretical properties provide a reference for the determination of $\sigma_\dg$ from the asymptotic point of view. 
\end{itemize}
}

\subsection{Theoretical Guarantees on the Algorithm}\label{sec:alg_theory}

We first introduce the necessary notations and conditions.
\begin{enumerate}[label=(C\arabic*), start=1]
	\item \label{con:bound-variance}For given $i\in \vertice$ and $\dg_{\max}$, define the random vector set
	$$
	\Lambda_{i,\dg}\coloneqq\Big\{\widehat{\intpar}_{i}:\, \widehat{\intpar}_{i} = \argmax_{\supp(\intpar_{i}) = \sactive}\npl_n(\intpar_{i}), \forall \sactive\text{ s.t. }|\sactive| \leq \dg\Big\}\bigcup\Big\{\intpar^*_{i}\Big\},
	$$
	there exists a universal constant $c\geq 1$ such that
	$$P\left\{\max_{i\in \vertice}\max_{\intpar_{i}\in \Lambda_{i,\dg_{\max}}}\|\intpar_{i}\|_1\leq c\ms\right\} < \varepsilon_{0} = o(1).$$
\end{enumerate}
\begin{remark}
	Condition~\ref{con:bound-variance} places a restriction on the magnitude of SLIDE across all possible models, since the curvature of PL may vanish at some points far away from the origin. In the context of generalized linear models, similar conditions with the same purpose are commonly encountered and are intended to avoid infinitely large or small variance of $y_i$, as well as ensure the existence of the Fisher information for statistical inference \citep{rigollet2012kullback, fan2013tuning}.
\end{remark}
For arbitrary support size $\dg$, we define 
$\mxeg_{\dg}^0 \coloneqq \max\limits_{\sactive:|\sactive| \leq \dg}\lambda_{\max}(\mathbb{E}[\data_{\sactive}\data_{\sactive}^{\top}])$, $\mneg_{\dg}^0\coloneqq \min\limits_{\sactive:|\sactive| \leq \dg}\lambda_{\min}(\mathbb{E}[\data_{\sactive}\data_{\sactive}^{\top}])$, and 
$\mxsg_{\dg}^0 \coloneqq \max\limits_{\sactive, \mathcal{B}: \substack{|\sactive| \leq \dg,\,|\mathcal{B}| \leq \dg, \sactive\cap\mathcal{B}=\varnothing}
}\sigma_{\max}(\mathbb{E}[\data_{\sactive}\data_{\mathcal{B}}^{\top}])$.  
\begin{enumerate}[label=(C\arabic*), start=2]
\item \label{con:technical} For some small constant $\gap >0$, denote $\mxeg_\dg = \mxeg_\dg^0 + \gap$, $\mnhs_\dg = e^{-2c\ms} (\mneg_\dg^0 - \gap)$ and $\mxsg_\dg = \mxsg_\dg^0 + \gap$, with constant $c$ as defined in \ref{con:bound-variance}. There exists some constant $\Delta\in (0, 1)$, such that $0\leq\tech({\dg_{\max}}, \ms) < 1$. Here 
$\tech(\dg, \ms) \coloneqq \frac{
    C_{1}(\dg, \ms)
}{
    (1-\Delta)C_{2}(\dg, \ms)
}$ and
\begin{align*}
    C_{1}(\dg, \ms) = \frac{2\mxsg_\dg^2}{\mnhs_\dg^2}
    \left(\frac{\mxeg_\dg}{2} + \frac{\mxsg_\dg^2}{\mnhs_\dg} + \frac{\mxeg_\dg\mxsg_\dg^2}{2\mnhs_\dg^2} \right)
    \left(3 + \frac{2\mxsg_\dg}{\mnhs_\dg}\right)^2, \;\; C_{2}(\dg, \ms) = \frac{\mnhs_\dg}{2} - \frac{\mxsg_\dg^2}{\mnhs_\dg} - \frac{\mxeg_\dg\mxsg_\dg^2}{2\mnhs_\dg^2}.
\end{align*}
\end{enumerate}
\begin{remark}
	Condition~\ref{con:technical} is a technical requirement. The quantities $\mxeg_\dg^0$, $\mneg_\dg^0$ and $\mxsg_\dg^0$ are widely used in modeling high-dimensional data. They restrict the correlation among a small number of variables and, in turn, guarantee the identifiability of the true active set. $\mxeg_\dg$ and $\mnhs_\dg$ are actually upper and lower bounds for the spectrum of the Hessian matrix over supports with a restricted cardinality. Condition~\ref{con:technical} is satisfied when they are closed enough and $\mxsg_\dg$ is sufficiently small. 
	Specifically, in an asymptotic point of view, in order to guarantee that $C_{2}(\dg, \ms)>0$, $\mxsg_{\dg}$ is required to decay exponentially with $\ms$. 
	Intuitively, this requires PL to be both smooth and convex enough, and column vectors of the data matrix are nearly orthogonal. In particular, this condition holds trivially if the data matrix has orthogonal columns.
\end{remark}
\begin{enumerate}[label=(C\arabic*), start=3]
	\item $\sigma_\dg = \Theta\left(\frac{\dg \log p \log\log n}{n}\right)$ for any $1\leq \dg \leq\dg_{\max}$. \label{con:threshold}
\end{enumerate}
\begin{remark}
	Conditions~\ref{con:threshold} can be decomposed into $\sigma_\dg = O\left(\frac{\dg \log p \log\log n}{n}\right)$ and $\sigma_\dg = \Omega\left(\frac{\dg \log p \log\log n}{n}\right)$. The former is designed to prevent $\sigma_\dg$ from becoming too large, allowing for necessary splicing steps, while the latter aims to reduce unnecessary iterations.
\end{remark}

We now present the statistical guarantee of Algorithm~\ref{alg:abess-mcl}. The following theorem asserts that the solution of Algorithm~\ref{alg:abess-mcl} consistently recovers the underlying Ising model with high probability. 
\begin{theorem}[Consistency of Algorithm~\ref{alg:abess-mcl} in Reconstructing Ising Models]\label{thm:Ising_gic}
Denote $\widehat{\intpar}$ as the output of Algorithm~\ref{alg:abess-mcl}. Assume Conditions~\ref{con:bound-variance}-\ref{con:threshold} hold and that $c_n \in [(1+\Delta)K_0 e^{4\ms}(1+\delta), \frac{C_2(\dg_{\max},\ms)\ws^2 n}{\dg_{\max}\log p}]$ for some constant $K_0>0$ and arbitrary constant $\delta>1$. Suppose for some universal constant $K_1>0$
\begin{equation}\label{eq:n_for_gic}
n >  \frac{K_1 \dg_{\max}\log p}{[1-\tech(\dg_{\max},\ms)](1-\Delta)C_2(\dg_{\max},\ms) \ws^2}\log\left[\frac{\dg_{\max}\log p}{
   [1-\tech(\dg_{\max},\ms)](1-\Delta)C_2(\dg_{\max},\ms) \ws^2}\right],
\end{equation} 
then the support set of $\intpar^*$ is selected with high probability: 
\begin{align*}
    P\left(\supp(\widehat{\intpar}) = \supp(\intpar^{*})\right) \geq 1- \varepsilon_{0} - \varepsilon_{1}(\dg_{\max}) -\varepsilon_{2}(\dg_{\max}) - \varepsilon_{3}(\dg_{\max}) - \varepsilon_{4}(\dg_{\max}) - \dg_{\max} p^{1-\delta}.
\end{align*}
Here 
$\ds\varepsilon_1(\dg) = 2p^{3}\exp\Big\{-\frac{n\gap^2}{2\dg^{2}}\Big\}$, $\ds \varepsilon_2(\dg) = 2p^2\exp\Big\{-\frac{n\ws^2 t^2(\dg, \ms)}{8\dg}\Big\}$, $\ds \varepsilon_3(\dg) = 2p^2\exp\Big\{-\frac{n e^{-4\ms}(\mneg_{\dg}^0 - \gap)^2}{72\dg^2 \mxeg_{\dg}^2}\Big\}$ and $\ds \varepsilon_4(\dg) = 2p^2\exp\Big\{-\Big( \frac{n \tilde{t}^2(\dg, \ms)}{8^3\dg^4} \Big)^{\frac{1}{3}}\Big\}$. The explicit forms of $t(\dg, \ms)$ and $\tilde{t}(\dg, \ms)$ are provided in the Supplementary Materials.
\end{theorem}
\begin{remark}
	Theorem~\ref{thm:Ising_gic} helps reveal the sample complexity of Algorithm~\ref{alg:abess-mcl} (denoted as $\algn$) with respect to $\ws$, $p$, $\dg$ and $\ms$. It is easy to see that $\algn \;\propto\; \ws^{-2} \log \ws^{-2}$ and $\algn \;\propto\; \log p \log\log p$. And thus, for $\ws$ and $p$, $\algn$ lies in the globally optimal region up to some logarithmic terms. The relationship between $\algn$ and $\dg$ is not directly evident, as it depends on two quantities $C_2(\dg, \ms)$ and $\nu(\dg, \ms)$\footnote{Here we set $d_{\max} = C \dg$, where $C > 1$ is a constant to simplify notations and make the results of Theorem~\ref{thm:Ising_gic} being comparable to Theorem~\ref{thm:sample_size_d}.}. By lower-bounding $C_2(\dg, \ms)$, upper-bounding $\nu(\dg, \ms)$, and assuming $\ws$ is known (as the baselines in Table~\ref{tab:sc-comparison2}), we have $\algn \;\propto\; d^6$. A similar argument for $\ms$ yields $\algn \;\propto\; e^{8(c+1)\ms}$ for some constant $c \geq 1$ defined in Condition~\ref{con:bound-variance}. Although there exists a gap between $\algn$ and $n^*$ in terms of $\ms$ and $\dg$, this gap appears to be an artifact of the worst-case analysis. Indeed, Figures~\ref{fig:degree-scale}--\ref{fig:temp-scale} show that the empirical sample complexity $\empn$ of Algorithm~\ref{alg:abess-mcl} matches—or even outperforms—the theoretical SLIDE estimator. For example, $\empn \;\propto\; e^{3.5\ms}$ grows more slowly than SLIDE's rate $e^{4\ms}$.
\end{remark}

\noindent Since the support set is consistently detected, we can easily establish the property of parameter estimation.
\begin{corol}\label{coro:coupling}
 Under the same conditions and notations in Theorem~\ref{thm:Ising_gic}, with additional probability of error $\varepsilon$, the (squared) $\ell_2$ error bound of the $i$-th column of $\widehat{\intpar}$ returned by Algorithm~\ref{alg:abess-mcl} satisfies:
\begin{align*}
    \| \widehat{\intpar}_i -  \intpar^*_i \|^2_2 \leq  \frac{1}{\mnhs^2_\dg} \frac{16\dg}{n}\log\left(\frac{2p}{\varepsilon}\right)
\end{align*}
for all $i =1,\ldots, p$. 
\end{corol} 

The next theorem claims that, in a high probability sense, Algorithm~\ref{alg:abess-mcl} has a polynomial-time complexity. 
\begin{theorem}\label{thm:complexity}
    Assume Conditions~\ref{con:bound-variance}-\ref{con:threshold} hold and Algorithm~\ref{alg:abess-mcl} successfully recovers the true graphical structure $\supp(\widehat{\intpar}) = \supp(\intpar^*)$ (by Theorem~\ref{thm:Ising_gic}, this event occurs with high probability for a sufficiently large sample size $n$). Then the computational complexity of Algorithm~\ref{alg:abess-mcl} for a given $\dg_{\max}\geq \max\limits_{i\in \vertice} |\supp(\intpar^*_{i})|$ is
\begin{align*}
    O\Big(\big(\frac{\dg_{\max}}{1-\tech(\dg_{\max},\ms)}\ln \frac{n\ms}{\min\limits_{i\in \vertice} |\supp(\intpar^*_{i})| \log p \log\log n } + \frac{n\ms}{ \log p \log\log n}\big)\big( pn\dg^2_{\max} + \dg_{\max}p^2 \big)\Big).
\end{align*}
\end{theorem}
We claim that the computational complexity of Algorithm~\ref{alg:abess-mcl} scales as a polynomial in $(n, p,\dg)$. In other words, under certain conditions, our algorithm solves the intractable problem of learning sparse Ising model in a polynomial time. To see this, we only need to verify that $\frac{1}{1- \tech(\dg,\ms)} = \frac{(1-\Delta) C_2}{(1-\Delta) C_2 - C_1}$ scales at most polynomially with $(n, p,\dg)$. This is true since $\lambda_{\max}(\mathbb{E}[\data_{\sactive}\data_{\sactive}^{\top}])\leq d$, $\sigma_{\max}(\mathbb{E}[\data_{\sactive}\data_{\mathcal{B}}^{\top}])\leq d$, and $\lambda_{\min}(\mathbb{E}[\data_{\sactive}\data_{\sactive}^{\top}])\geq \frac{e^{-2\ms}}{\dg + 1}$ where the last inequality is also presented in Lemma~7 of \citet{vuffray2016interaction}. Note that the complexity grows exponentially fast as $\ms$ increases, which is aligned with our expectation since the complexity is derived for the case that our algorithm successfully recovers the graphical structure, which, in turn, requires a sample size $n$ to scale exponentially with $\ms$. 

\section{Simulation Studies}\label{sec:simulation}
The simulation studies are divided into three part. The first part would like to certify SLIDE has a superiority on empirical sample complexity $\empn$ with respect to $\dg$ and $\ms$ (see Section~\ref{sec:numeric-complexity}). And the second part presented in Section~\ref{sec:numeric-hd-regime} is mainly interested in another practice scenario where the number of parameters are comparable or even larger than $n$, so called high-dimensional regime. The section is closed with numerical experiments on computation performance. 

\subsection{Numerical Advantages on Sample Complexity}\label{sec:numeric-complexity}
In this part, we leverage the empirical sample complexity $\empn$ to evaluate and compare methods. Roughly speaking, $\empn$ is empirical counterpart of $n^*$, recording the minimal numerical sample size that exactly recovers the underlying graphical structure. Its detailed definition is deferred to the Supplementary Materials. We also empirically verify $\empn$ is proportional to $\log p$ and $\ws^{-2}$, which is matched to the sample complexity of SLIDE. Please see the Supplementary Materials for detailed results.

\subsubsection{Experiments on $\dg$}\label{sec:experiment-on-d}
We first investigate $\empn$ with respect to $\dg$ to illustrate the efficacy of SLIDE. The random regular graph (RRG) topology is used in this investigation because it allows the generation of a network of any degree $\dg < p$ when $p, \ws, \ms$ are all fixed. To be more specific, to disentangle the effects of $d$ from $\ws$ and $\ms$, we force each node having exactly one neighbor with interaction $\ws$ and the other interaction being $\beta$ such that $\ms = (\dg - 1) \beta + \ws$ is a fixed constant. 

\begin{figure}[htbp]
    \begin{center}
        \includegraphics[width=0.8\linewidth]{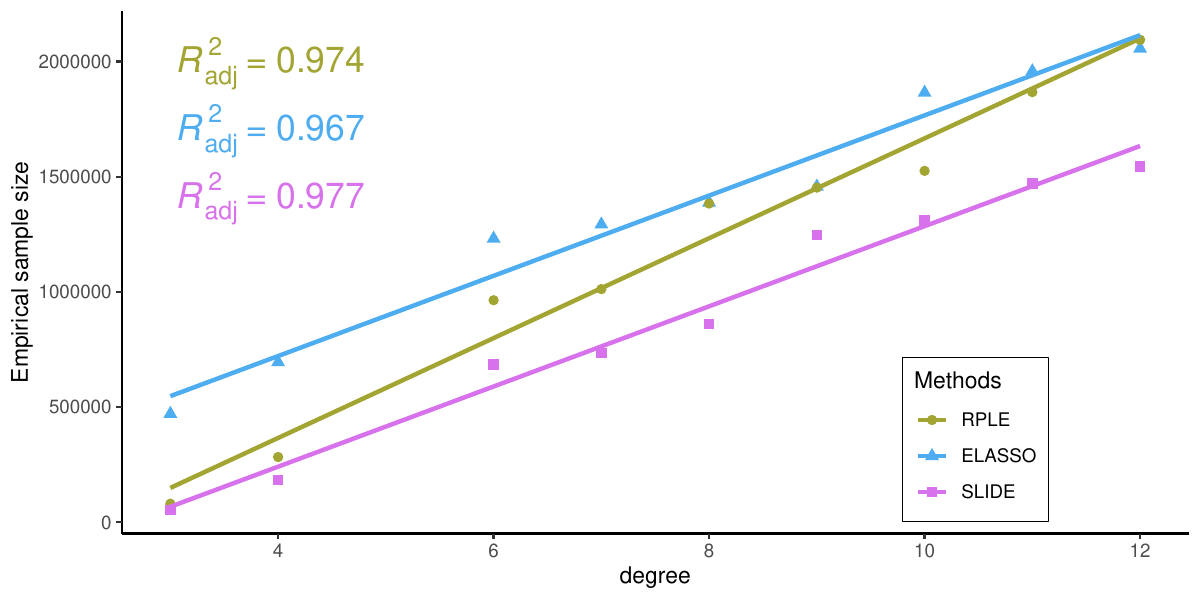}
    \end{center}
    \vspace{-20pt}
    \caption{The maximum degree $\dg$ ($x$-axis) versus empirical sample size ($y$-axis) scatterplot under the ferromagnetic RRGs. {The three straight lines} are characterized by equations $y = a + bx$, where the coefficients $a, b$ are estimated by linear regression. The adjusted $R^2$ of each fitted linear model is reported in the left-top corner.}\label{fig:degree-scale}
\end{figure}

Figure~\ref{fig:degree-scale} provides a graphical representation of the corresponding Ising model in its lower right corner. A linear relationship between $\empn$ and $\dg$ appears to exist in SLIDE (see Figure~\ref{fig:degree-scale}), with adjusted $R^2$ reaches 0.97. This supports the $O(\dg)$ lower bound on $\dg$ obtained from our theoretical analysis of the SLIDE estimator. {For the sake of comparison, we also include two benchmarked methods: (i) LASSO-\underline{r}egularized \underline{p}seudo-\underline{l}ikelihood \underline{e}stimator (RPLE) and (ii) extended LASSO \citep[ELASSO,][]{van2014new}. As can be seen in Figure~\ref{fig:degree-scale}, the slope of the purple line is less steep than that of the brown line, indicating that our approach is more sample-efficient than RPLE, as the rate of growth of the empirical sample size for SLIDE is lower than that for RPLE. Besides, the slope of the blue line is less steep than that of the brown line, implying ELASSO does improve PRLE in terms of sample efficiency. However, ELASSO still requires more empirical samples than SLIDE. In the Supplementary Materials, we provide an additional benchmarked Ising model to show the sample complexity of RPLE/ELASSO does not grow linearly with respect to $\dg$ while that of SLIDE does. This implies the dominant advantage of SLIDE over RPLE/ELASSO.} 

\subsubsection{Experiments on $\ms$}\label{sec:experiment-on-ms}

Here, we empirically achieve the scale of $\empn$ w.r.t. $\ms$ across a variety of graphical topologies and interaction patterns via a large set of numerical tests. We take into account five benchmarked distinct Ising models considered in \citet{lokhov2018optimal}. These Ising models cover two reference topologies including RRGs and a periodic-boundary square lattice (PBSL). The two topologies avoid the instability caused by the heterogeneity of nodes to make it simpler to extract the proper scaling with regard to $\ms$. The two sorts of interactions are considered: (i) $\intpar_{ij} > 0$ and (ii) $\intpar_{ij} \in \mathbb{R}$, where all interaction magnitudes $| \intpar_{ij} |$ are set to a positive value $\beta$ except for the weakest signal $\ws$. It is necessary to fix one coupling in cases (i) to $\ws$ and $-\ws$, respectively, and fix two couplings in cases (ii) to $\ws$ and $-\ws$ in order to separate the effects of $\ws$ and $\ms$. The combination of two topologies and two interaction types result in four Ising models. The remain one Ising model is a PBSL with interaction (i) but one of interaction is weakly negative, which is considered as a hard model for learning graphical structure \citep{lokhov2018optimal}.  

\begin{figure}[htbp]
	\centering
	\vspace*{-33pt}
	\includegraphics[width=1.0\textwidth]{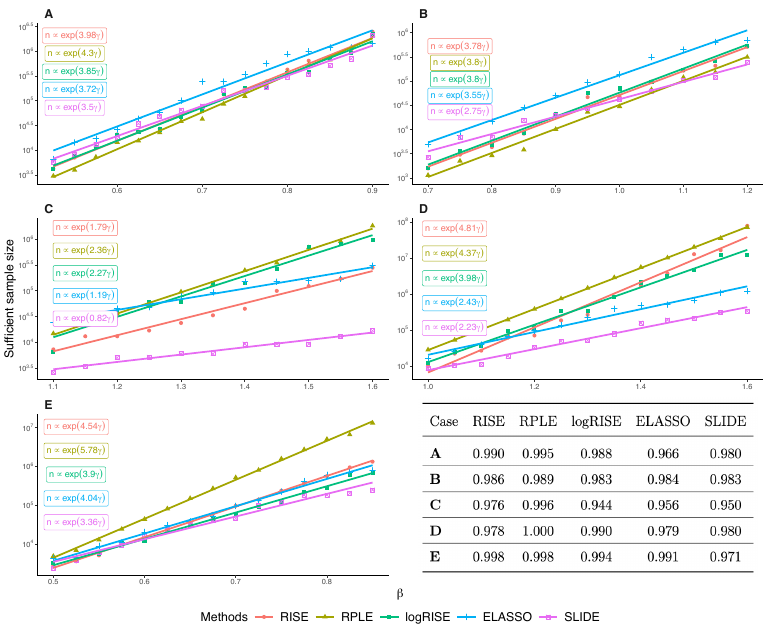}
	\vspace{-45pt}
	\caption{Interactions $\beta$ ($x$-axis) versus empirical sample size ($\log_{10}$-transformed $y$-axis) scatterplot in five types of Ising models. A method with the smallest slope is the best one; otherwise, the worst. The left panels and right panels correspond to RRGs and PBSLs. The scatterplots in upper, middle, and bottom panels correspond to interaction types (i), (ii), and (iii), respectively. {The top-left corner of each panel presents the exponent scaling $b$ of five methods, which are given by estimating the coefficients of linear models $\log(\empn) = a + b (\dg \beta)$, where $d = 3$ for RRGs and $d = 4$ for PBSLs. The table in the right-bottom corner reports the adjusted $R^2$ of the fitted linear models.}}\label{fig:temp-scale}
\end{figure}

For comparison, a \underline{r}egularized \underline{i}nteraction \underline{s}creening \underline{e}stimators (RISE) and its variant, logRISE, are also included in this analysis, where logRISE is a $\ms$-optimal estimator \citep{lokhov2018optimal}. {Besides, the benchmarked methods in Section~\ref{sec:experiment-on-d}, RPLE and ELASSO, are also taken into account.} From Figure~\ref{fig:temp-scale}, the $n_\textup{emp}$ of {five methods} grows exponentially with regard to $\ms$ (as evidence by large adjusted $R^2$ values). It's clear that the SLIDE has the finest scaling property with respect to~$\ms$ because its growth rate is the lowest of all these approaches across all five Ising models. Even in its worst case, the empirical scaling of SLIDE (i.e., $\exp\{3.5\}$ on PBSL with interaction (i)) is still in the optimal range with respect to the information-theoretic prediction. According to \citet{lokhov2018optimal}, a scaling factor of 3.9 for logRISE places it in the $\ms$-optimal regime, while the RISE Ising model estimator comes in at {number four among the five methods} under comparison. {Finally, ELASSO outperforms RPLE in terms of scaling with respect to $\ms$, likely due to the effectiveness of EBIC in identifying the true network structure as the sample size increases \citep{barber2015high}.}

\subsection{High-dimensional Regime}\label{sec:numeric-hd-regime}

To compare the estimation accuracy of different methods in this regime, we consider metrics from two aspects, one focuses on structure recovery and the other on parameter estimation. We measure the accuracy of structure recovery by true positive rate (TPR), false positive rate (FPR) and Matthews correlation coefficient (MCC), which are defined as
\begin{align*}
  \textup{TPR} &= \frac{\textup{TP}}
  {\textup{TP} + \textup{FN}}, \; 
  \textup{FPR} = \frac{\textup{FP}}
  {\textup{TN} + \textup{FP}}, \\
  \textup{MCC} &= \frac{\textup{TP} \times \textup{TN} - \textup{FP} \times \textup{FN}}
  {\sqrt{(\textup{TP} + \textup{FP})(\textup{TP} + \textup{FN})(\textup{TN} + \textup{FP})(\textup{TN} + \textup{FN})}},
\end{align*}
where TP, TN, FP and FN refer to
\begin{align*}
  \textup{TP} &= \left|\left\{ (i, j) \mid \widehat{\intpar}_{ij} \neq 0 \cap \intpar^{*}_{ij} \neq 0 \right\}\right|,
  \textup{TN} = \left|\left\{ (i, j) \mid \widehat{\intpar}_{ij} = 0 \cap \intpar^{*}_{ij} = 0 \right\}\right|,\\
  \textup{FP} &= \left|\left\{ (i, j) \mid \widehat{\intpar}_{ij} \neq 0 \cap \intpar^{*}_{ij} = 0 \right\}\right|,
  \textup{FN} = \left|\left\{ (i, j) \mid \widehat{\intpar}_{ij} = 0 \cap \intpar^{*}_{ij} \neq 0 \right\}\right|.
\end{align*}
In terms of interaction estimations, mean square error (MSE) between the estimated and truth is a widely adopted measure 
for the assessment of parameter accuracy. 
It is defined as $\textup{MSE}(\intpar, \widehat{\intpar}) = \frac{2}{p(p-1)}\sum\limits_{i < j} (\widehat{\intpar}_{ij} - \intpar_{ij})^2$. 

\begin{figure}[t]
	\begin{center}
		\includegraphics[width=1.0\linewidth]{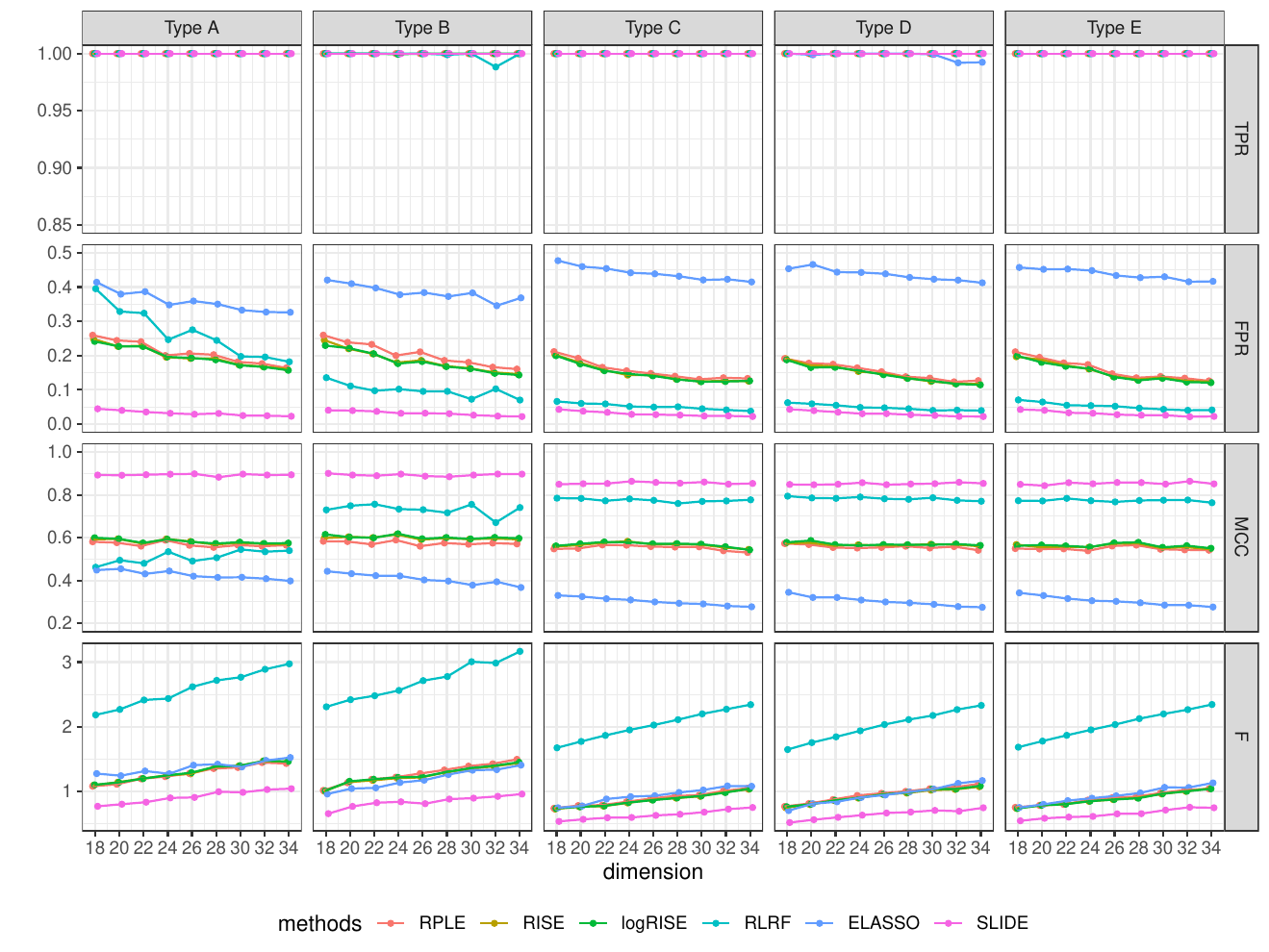}
	\end{center}
	\vspace*{-33pt}
	\caption{Structure recovery and parameter estimation performance comparison of the competing methods under high-dimensional settings.}\label{fig:high-vs}
\end{figure}

For each underlying graph presented in Section~\ref{sec:experiment-on-ms}, we generate datasets with 200 observations where the dimension of binary variables varies from 18 to 34. Then, we apply our methods and the competitor on these datasets according to detailed settings shown in the Supplementary Materials. {Here, we also include one additional benchmarked method that leverages an upper bound on the log-partition function to simplify computation, which is referred to RLPF\footnote{RLPF is not considered in Section~\ref{sec:numeric-complexity} as it cannot exactly recover structure of Ising model when sample size is sufficiently large (see the Supplementary Materials).}.} The simulation results based on 100 random replications are demonstrated in Figure~\ref{fig:high-vs}. From the exhibited comparison, we can see that, compared to the state-of-the-art methods with $\ell_1$ regularization, SLIDE has a very competitive TPR, but it also would have a significantly lower FPR, which leads to its MCC, a measure that shows overall structure learning performance, would be higher than that of other approaches. Thus, from the perspective of graph support recovery, SLIDE outperforms other methods under the high-dimensional setting. The high false inclusion is not surprising because the biases caused by the $\ell_1$ penalty force the models to include more edges for compensation \citep{fan2001variable,xue2012nonconcave}. 

On the other hand, in terms of MSE, it is obvious that MSE generally increases as the dimension gets higher, whatever method is implemented, and the result may be due to a faster increase in MSE than in dimension. What's more, SLIDE also owns the best performance in parameter estimation. The outstanding parameter estimation comes from the unbiasedness of the $\ell_0$ constraint that only encourages sparsity but does not shrink the estimation. Therefore, we can conclude that SLIDE owns outstanding performance and can provide a good estimator in the high-dimensional regime. 

\subsection{Runtime Analysis}\label{sec:experiment-runtime}
Adopting the numerical settings in Section~\ref{sec:numeric-hd-regime}, we further investigate the computational performance in these cases. As shown in the Supplementary Materials, among all nodewise methods, the SLIDE demonstrates the fastest computational speed on five benchmarked Ising models, followed by ELASSO, logRISE, RISE, and finally RPLE method. Remarkably, the computational advantage is tremendous, with up to a 100-fold speedup compared to RPLE/RISE/logRISE and 10-fold speedup compared to ELASSO. The results in the Supplementary Materials refute the widely held belief that the best-subset selection approach is not feasible for learning high-dimensional Ising models. More importantly, the fitted curves (displayed in the top-left corner of each panel) depicting dimensionality against computation time show that the runtime of SLIDE scales quadratically with respect to $p$. This confirms the computational complexity of SLIDE is polynomial. {Finally, RLPF is as competitive as SLIDE when $p$ is small, and it marginally outperforms SLIDE when $p$ is large. However, this slight computational advantage of RLPF is not adequately attractive, as RLPF cannot exactly reconstruct some Ising models.}

\section{US Senate Voting Data Analysis}

How humans interact with each other is one of the most important problems in social science \citep{shneidermanScience2008, lazerComputationalSocialScience2009}. In the current and extremely divisive political environment in the U.S. and the world, it is more important than ever to study how individuals, including politicians, behave; for example, how politicians vote \citep{shneidermanScience2008}, and why bipartisanship occurs when it is extremely unpopular and politically risky within the base of their party. Studying social interactions, however, is challenging because we cannot see people's inner thoughts. Reconstructing the hidden information from observable individual behaviors, such as politicians' voting records, is thus critically vital but also highly difficult. Ising model reconstruction is a fundamental tool in sociology for this purpose. 

\begin{figure}[t]
    \centering
    \includegraphics[width=0.81\textwidth]{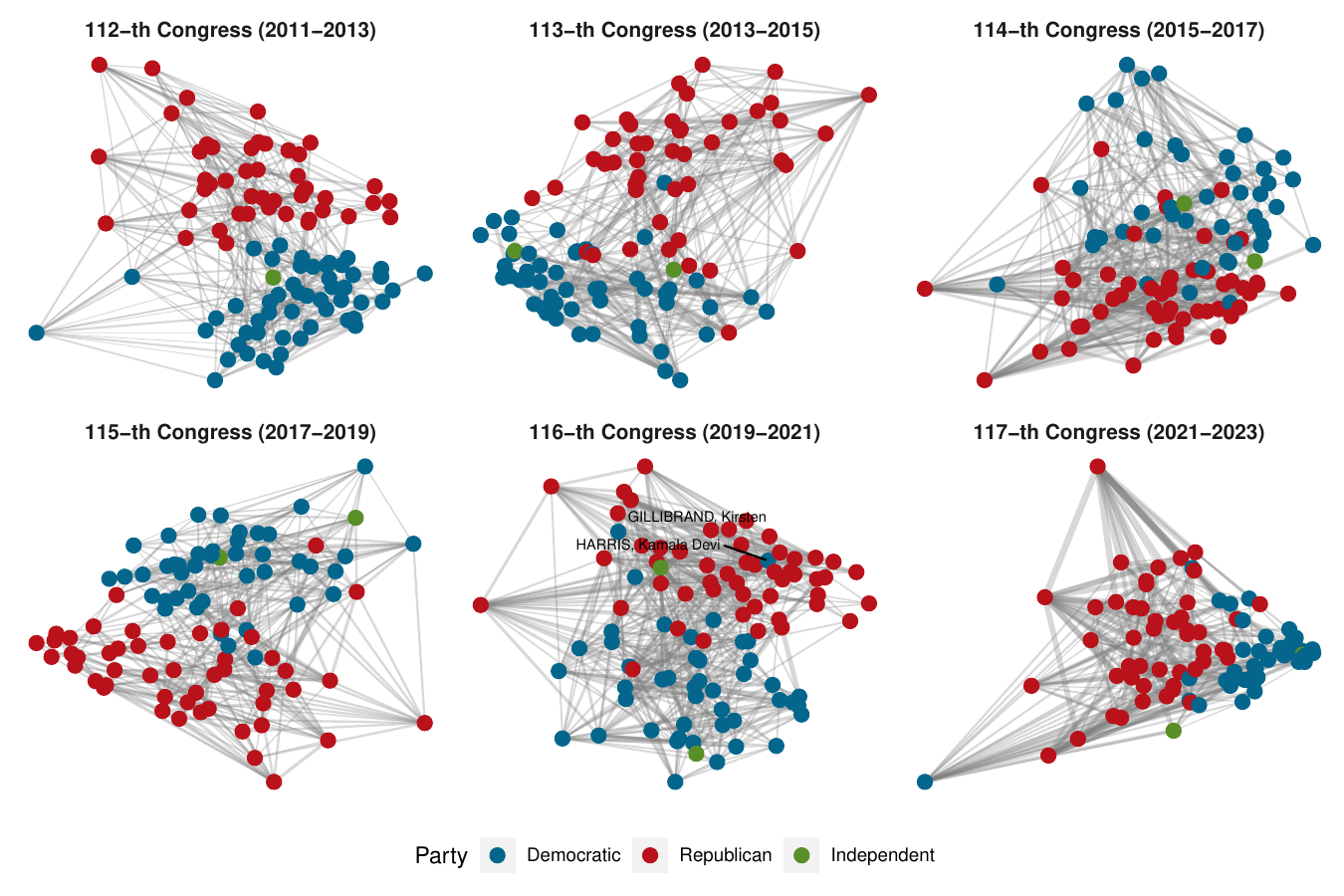}
    \vspace*{-18pt}
    \caption{Voting records from the 112th to 117th Congresses are estimated by the SLIDE  for the US Senate. The SLIDE uses the first two eigenvectors of the estimated interaction matrix to determine the node layout. 
    The numbers of senators in the six Congresses are: 99, 98, 100, 96, 98, 97, respectively. 
    Only senators who have neighbors are shown.
   }
    \label{fig:senate_voting_all}
\end{figure}

The US Senate Congress Voting Dataset is a popular test case for Ising model reconstruction. We consider data from the $112$th to the $117$th Congresses that are fetched from the VoteView\footnote{\url{https://voteview.com/}}. The Senate votes ``Yea'' or ``Nay'' on federal legislation that affects all aspects of US domestic and foreign policy. If a vote is missing, it is imputed as ``Nay'' \citep{banerjee2008model}. Binary variables can be used to describe senators' votes. For each Congress, an Ising model can be built to shed light on senators' interactions with one another (see Figure~\ref{fig:senate_voting_all}). The senators can be loosely classified into two groups in this figure: the Republicans and the Democrats. In line with both common sense and prior research of \citet{banerjee2008model} in just one Congress, it stands to reason that those within a party will have a stronger bond with one another than those from opposing parties.
From Figure~\ref{fig:senate_voting_all}, the 112th and 117th Congresses were visually more distinct. 
The 112th Congress had a greater degree of partisanship than any other. 
The Washington Post\footnote{\url{https://www.washingtonpost.com/news/wonk/wp/2013/01/17/its-official-the-112th-congress-was-the-most-polarized-ever/}} reported that the 112th Congress was the most polarized ever as measured by the DW-NOMINATE score \citep{poole1985spatial}. 
We apply the spectral clustering method \citep{shi2000normalized} on the reconstructed Ising model to make predictions, and get an unsupervised classifier with an accuracy rate of nearly $90\%$. 
Next, the Democratic senators in the 117th Congress displayed the most concentrated and out-of-the-ordinary pattern. 
This Senate is made up of 50 Republicans, 48 Democrats, and 2 Independents who caucus with the Democrats. 
This makes the Senate effectively evenly split, only the third time in US history. 
Without any leeway for dissent, to support the agendas of the Democrat president, the Democrats need to be fully united\footnote{\url{https://fivethirtyeight.com/features/why-house-democrats-may-be-more-united-than-they-seem/}}.

On a micro level, the political intention of the senators can be investigated from estimated networks and important insights can be unearthed. For example, in the 116th Congress, the voting pattern of two Democrat senators, Kamala Harris and Kirsten Gillibrand, had voted with the Republican Party. Although there is no particularly media attention for the fact that the current vice president sometimes agreed with the Republicans, her fellow senator Gillibrand\footnote{\url{https://twitter.com/sengillibrand/status/1425567127696777216}}, who worked together with her during the 2019 Democratic presidential primary debate\footnote{\url{https://www.refinery29.com/en-us/2019/08/239485/kamala-harris-kirsten-gillibrand-democratic-debate-joe-biden-op-ed}}, was known to once hold conservative political views\footnote{\url{https://edition.cnn.com/2019/01/20/politics/gillibrand-conservative-past-iowa-democrats}}. The estimated interaction of 0.57 between these leaders may suggest that the vice president might be more conservative than the public is informed. 

\section{Conclusion and Discussion}\label{sec:conclusion-discussion}
Finding the optimal sample complexity solution for the reconstruction of the Ising model is an open problem that is explored in this study. We introduce the SLIDE approach, and from theory and computation, shows that it is provable global. Moreover, it can be solved with polynomial complexity under realistic conditions. To the best of our knowledge, it is the first known approach that reaches the global optimality in a polynomial time. This means that our method can reconstruct various Ising models, even challenging models with low temperature and hubs, where existing methods require a larger sample complexity. Evidence from the real-world example also suggests that the SLIDE can recover the interactions of complex systems and shed light on subtle information from apparently opaque observations. 

More efforts are deserves to enlarge the impact of sparse learning approach and the algorithm for solving. For example, one promising direction is extending our approach to recover Potts models for discrete variables \citep{wuPottsModel1982} and {Ising model with random effects \citep{fan2012variable, he2023hidden_a, he2023hidden_b}}. From application, our solution facilitates the new findings on cutting-edge scientific problems such as functional connectome and disease connectome.

\bibliographystyle{agsm}
\bibliography{reference.bib}

\end{document}